\documentstyle[12pt]{article}

\input epsf.tex

\def\ll{\lambda\lambda}
\def\vev#1{\mathord < #1 \mathord >}

\def\Sb{\bar{S}}
\def\Tb{\bar{T}}
\def\Yb{\bar{Y}}

\def\beq{\begin{equation}}
\def\eeq{\end{equation}}
\def\TR{T+\Tb-Y\Yb}

\begin{document}

\begin{flushright}
   hep-th/9508173\\
   TUM-HEP 226/95\\
   SFB 375/20\\
\end{flushright} \vspace{1ex}

\begin{center}{ \large \bf

  Gaugino Condensation and the Vacuum Expectation\\
Value of the Dilaton$^\ast$\\
}

\vspace{1cm}

{\bf A. Niemeyer$^\dagger$ and
H. P. Nilles$^{\dagger,\ddagger}$}

\vspace{0.3cm}
 $^\dagger$ {Physik Department} \\
       {\em Technische Universit\"at M\"unchen} \\
       {\em D--85748 Garching, Germany}

\vspace{0.3cm}
 $^\ddagger$ Max Planck Institut f\"ur Physik \\
  {\em Heisenberg Institut}\\
  {\em D--80805 Munich, Germany}

\vspace{0.5cm}

ABSTRACT
\end{center}

\noindent The mechanism of gaugino condensation has emerged as a prime
candidate for supersymmetry breakdown in low energy effective supergravity
(string) models. One of the open questions in this approach concerns the size
of the gauge coupling constant which is dynamically fixed through the vev of
the dilaton. We argue that a nontrivial gauge kinetic function $f(S)$ could
solve the potential problem of a runaway dilaton. The actual form of $f(S)$
might be constrained by symmetry arguments.

\vspace{5cm}

\noindent \begin{tabular}{c}
\hline \hspace*{15cm}
\end{tabular}

\noindent $^\ast$ Invited talk at the International Workshop on Supersymmetry
and Unification of Fundamental Interactions, SUSY 95, Ecole Polytechnique,
Palaiseau, France, May 15-19, 1995

\newpage

One of the prime motivations to consider the supersymmetric extension of the
standard model is the stability of the weak scale ($M_W$) of order of a TeV in
the presence of larger mass scales like a GUT-scale of $M_X=10^{16}\;GeV$ or
the Planck scale $M_{Pl} \approx 10^{18}\;GeV$. The size of the weak scale is
directly related to the breakdown scale of supersymmetry, and a satisfactory
mechanism of supersymmetry breakdown should explain the smallness of
$M_W/M_{Pl}$ in a natural way. One such mechanism is based on the dynamical
formation of gaugino condensates that has attracted much attention since its
original proposal for a spontaneous breakdown of supergravity \cite{1}\cite{2}.
In this talk we shall address some open questions concerning this mechanism in
the framework of low energy effective superstring theories. This work has been
done in collaboration with Z. Lalak and appeared in ref. \cite{3}\cite{4}.

Before addressing these detailed questions let us remind you of the basic facts
of this mechanism. For simplicity we shall consider here a pure supersymmetric
($N=1$) Yang-Mills theory, with the vector multiplet $(A_\mu, \lambda)$
containing gauge bosons and gauge fermions in the adjoint representation of
the nonabelian gauge group. Such a theory is asymptotically free and we would
therefore (in analogy to QCD) expect confinement and gaugino condensation at
low energies \cite{5}. We are then faced with the question whether such a
nontrivial gaugino condensate $\mathord < \lambda\lambda \mathord > \neq 0$
leads to a breakdown of supersymmetry. A first look at the
SUSY-transformation on the composite fermion $\lambda\sigma^\mu A_\mu$
\cite{dfs}

\beq
\{Q,  \lambda\sigma^\mu A_\mu\}=\lambda\lambda + \ldots
\eeq

\noindent might suggest a positive answer, but a careful inspection of the
multiplet
structure and gauge invariance leads to the opposite conclusion. The bilinear
$\lambda\lambda$ has to be interpreted as the lowest component of the chiral
superfield
$W^\alpha W_\alpha=(\ll, \ldots)$ and therefore a non-vanishing vev of $\ll$
does not break SUSY \cite{6}. This suggestion is supported by index-arguments
\cite{7} and an effective Lagrangian approach \cite{8}. We are thus lead to the
conclusion that in such theories gaugino condensates form, but do not break
global (rigid) supersymmetry.

Not all is lost, however, since we are primarily interested in models with
local supersymmetry including gravitational interactions. The weak
gravitational force should not interfere with the formation of the condensate;
we therefore still assume  $\mathord < \lambda\lambda \mathord > = \Lambda^3
\neq 0$. This expectation is confirmed by the explicit consideration of the
effective Lagrangian of ref. \cite{1} in the now locally supersymmetric
framework. We here consider a composite chiral superfield $U=(u, \psi, F_u)$
with $u= \mathord < \lambda\lambda \mathord >$. In this toy model
\cite{1}\cite{2} we obtain the surprising result that not only $\vev u =
\Lambda^3 \neq 0$ but also $\vev{F_u} \neq 0$, a signal for supersymmetry
breakdown. In fact

\beq
\vev{F_u} = M_S^2 = \frac{\Lambda^3}{M_{Pl}},
\eeq

\noindent consistent with our previous result that in the global limit $M_{Pl}
\rightarrow \infty$ (rigid) supersymmetry is restored. For a hidden sector
supergravity model we would choose $M_S \approx 10^{11}\;GeV$ \cite{2}.

Still more information can be obtained by consulting the general supergravity
Lagrangian of elementary fields determined by the K\"ahler potential $K(\Phi_i,
{\Phi^j}^\ast)$, the superpotential $W(\Phi_i)$ and the gauge kinetic function
$f(\Phi_i)$ for a set of chiral superfields $\Phi_i=(\phi_i, \psi_i, F_i)$.
Non-vanishing vevs of the auxiliary fields $F_i$ would signal a breakdown of
supersymmetry. In standard supergravity notation these fields are given by

\beq
F_i = \exp(G/2) (G^{-1})^j_i G_j + \frac 14 \frac{\partial f}{\partial \Phi_k}
(G^{-1})^k_i \ll + \ldots ,\label{eq3}
\eeq

\noindent where the gaugino bilinear appears in the second term \cite{9}. This
confirms
our previous argument that $\vev \ll \neq 0$ leads to a breakdown of
supersymmetry, however, we obtain a new condition: $\partial f/\partial \Phi_i$
has to be nonzero, i.e. the gauge kinetic function $f(\Phi_i)$ has to be
nontrivial. In the fundamental action $f(\Phi_i)$ multiplies $W_\alpha
W^\alpha$ which in components leads to a form $\mbox{Re} f(\phi_i) F_{\mu\nu}
F^{\mu\nu}$ and tells us that the gauge coupling is field dependent. For
simplicity we consider here one modulus field $M$ with

\beq
\vev {\mbox{Re} f(M)} = 1/g^2.
\eeq

This dependence of $f$ on the modulus $M$ is very crucial for SUSY breakdown
via gaugino condensation. $\partial f/\partial M \neq 0$ leads to $F_M\approx
\Lambda^3/M_{Pl}$ consistent with previous considerations. The goldstino is the
fermion in the $f(M)$ supermultiplet. In the full description of the theory it
might mix with a composite field, but the inclusion of the composite fields
should not alter the qualitative behaviour discussed here. An understanding of
the mechanism of SUSY breakdown via gaugino condensation is intimately related
to the question of a dynamical determination of the gauge coupling constant as
the vev of a modulus field. We would hope that in a more complete theory such
questions could be clarified in detail.

One candidate of such a theory is the $E_8 \times E_8$ heterotic string. The
second $E_8$ (or a subgroup thereof) could serve as the hidden sector gauge
group and it was soon found \cite{10} that there we have nontrivial $f=S$ where
$S$ represents the dilaton superfield. The heterotic string thus contains all
the necessary ingredients for a successful implementation of the mechanism of
SUSY breakdown via gaugino condensation \cite{11}\cite{12}. Also the question
of the dynamical determination of the gauge coupling constant can be addressed.
A simple reduction and truncation from the $d=10$ theory leads to the following
scalar potential \cite{13}

\beq
V=\frac 1{16 S_R T_R^3} \left [ |W(\Phi) + 2 (S_R T_R)^{3/2} (\ll)|^2 + \frac
{T_R}3 \left |\frac{\partial W}{\partial \Phi} \right |^2 \right]\label{eq5},
\eeq

\noindent where $S_R=\mbox{Re} S$, $T_R=\mbox{Re} T$ is the modulus
corresponding
to the overall radius of compactification and $W(\Phi)$ is the superpotential
depending on the matter fields $\Phi$. The gaugino bilinear appears via the
second term in the auxiliary fields (\ref{eq3}). To make contact with the
dilaton field, observe that $\vev \ll = \Lambda^3$ where $\Lambda$ is the
renormalization group invariant scale of the nonabelian gauge theory under
consideration. In the one-loop approximation

\beq
\Lambda = \mu \exp \left ( -\frac 1{bg^2(\mu)}\right),
\eeq

\noindent with an arbitrary scale $\mu$ and the $\beta$-function coefficient
$b$. This then suggests

\beq
\ll \approx e^{-f} = e^{-S} \label{eq7}
\eeq

\noindent as the leading contribution (for weak coupling) for the functional
$f$-dependence of the gaugino bilinear\footnote{ Relation (\ref{eq7}) is of
course not exact. For different implementations see \cite{12}, \cite{14},
\cite{15}. The qualitative behaviour of the potential remains unchanged.}.

In the potential (\ref{eq5}) we can then insert (\ref{eq7}) and determine the
minimum. In our simple model (with $\partial W/\partial T=0$) we have a
positive definite potential with vacuum energy $E_{vac}=0$. Suppose now for the
moment that $\vev{W(\Phi)} \neq 0$\footnote{In many places in the literature it
is quoted incorrectly that  $\vev{W(\Phi)}$ is quantized in units of the Planck
length since $W$ comes from $H$, the field strength of the antisymmetric tensor
field $B$ and $H=\mbox{d}B -\omega_{3Y} + \omega_{3L}$ ($\omega$ being the
Chern-Simons form) \cite{15}. Quantization is
expected for $\vev{\mbox{d}B}$ but not for $H$.}.

\noindent $S$ will now adjust its vev in such a way that $|W(\Phi) + 2 (S_R
T_R)^{3/2} (\ll)| =0$, thus

\beq
|W(\Phi) + 2 (S_R T_R)^{3/2} \exp(-S)| = 0.
\eeq

\noindent This then leads to
broken SUSY with $E_{vac} =0$ and a fixed value of the gauge coupling constant
$g^2 \approx \vev{\mbox{Re} S}^{-1}$. For the vevs of the auxiliary fields we
obtain $F_S=0$ and $F_T\neq 0$ with important consequences for the pattern of
the soft SUSY breaking terms in phenomenologically oriented models \cite{brig},
which we shall not discuss here in detail.

Thus a satisfactory picture seems to emerge. However, we have just discussed a
simplified example. In general we would expect also that the superpotential
depends on the moduli, $\partial W/\partial T\neq 0$ and, including this
dependence, the modified potential would no longer be positive definite and one
would have  $E_{vac}<0$.

But even in the simple case we have a further vacuum degeneracy. For any value
of $W(\Phi)$ we obtain a minimum with $E_{vac}=0$, including $W(\Phi)=0$. In
the latter case this would correspond to $\vev \ll=0$ and $S\rightarrow
\infty$. This is the potential problem of the runaway dilaton. The simple model
above does not exclude such a possibility. In fact this problem of the runaway
dilaton does not seem just to be a problem of the toy model, but more
general. One attempt to avoid this problem was the consideration of several
gaugino
condensates \cite{kraslal}, but it still seems very difficult to produce
satisfactory potentials that lead to a dynamical determination of the dilaton
for reasonable values of $\vev S$. In some cases it even seems impossible to
fine tune the cosmological constant to zero. In absence of a completely
satisfactory model it is then also difficult to investigate the detailed
phenomenological properties of the approach. Here it would be of interest to
know the actual size of the vevs of the auxiliary fields $\vev{F_S}$,
$\vev{F_T}$ and $\vev{F_U}$. In the models discussed so far one usually finds
$\vev{F_T}$ to be the dominant term, but it still remains a question whether
this is true in general.

In any case it seems that we need some new ingredient before we can understand
the mechanism completely. It is our belief, that the resolution of all these
problems comes with a better understanding of the form of the gauge kinetic
function $f$ \cite{3}\cite{4}. In all the previous considerations one assumed
$f=S$. How general is this relation? Certainly we know that in one loop
perturbation theory $S$ mixes with $T$ \cite{stmix}, but this is not
relevant for our discussion and, for simplicity, we shall ignore that for the
moment. The formal relation between $f$ and the condensate is given through
$\Lambda^3 \approx e^{-f}$ and we have $f=S$ in the weak coupling limit of
string theory. In fact this argument tells us only that

\beq
\lim_{S\rightarrow \infty} f(S) = S\label{eq8}.
\eeq

Nonperturbative effects could lead to the situation that $f$ is a very
complicated function of $S$. In fact a satisfactory incorporation of gaugino
condensates in the framework of string theory might very well lead to such a
complication. In our work \cite{3} we suggested that a nontrivial $f$-function
is the key ingredient to better understand the mechanism of gaugino
condensation. We still assume (\ref{eq8}) to make contact with perturbation
theory. How do we then control $e^{-f}$ as a function of $S$? In absence of a
determination of $f(S)$ by a direct calculation one might use symmetry
arguments to make some progress. Let us here consider the presence of a
symmetry called $S$-duality which in its simplest form is given by a $SL(2,Z)$
generated by the transformations

\beq
S\rightarrow S+i,\quad S\rightarrow -1/S.
\eeq

Such a symmetry might be realized in two basically distinct ways:
the gauge sector could close under the transformation (type I) or being mapped
to an additional `magnetic sector' with inverted coupling constant (type II).
In the second
case one would speak of strong-weak coupling duality, just as in the case of
electric-magnetic duality \cite{seiwi}. Within the class of theories of type I,
however, we
could have the situation that the $f$-function is itself invariant under
$S$-duality; i.e. $S\rightarrow -1/S$ does not invert the coupling constant
since the gauge coupling constant is not given by $\mbox{Re} S$ but $1/g^2
\approx \mbox{Re} f$. In view of (\ref{eq8}) we would call such a symmetry
weak-weak coupling duality. The behaviour of the gauge coupling constant as a
function of $S$ is shown in Fig. 1. Our assumption (\ref{eq8}) implies that
$g^2 \rightarrow 0$ as $\mbox{Re} S \rightarrow \infty$ and by $S$-duality
$g^2$ also vanishes for $S\rightarrow 0$, with a maximum somewhere in the
vicinity of the self-dual point $S=1$. Observe that $S\approx 1$ in this
situation does not necessarily imply strong coupling, because  $g^2
\approx 1/\mbox{Re} f$ and even for $S\approx 1$, $\mbox{Re} f$ could be large
and $g^2<<1$, with perturbation theory valid in the whole range of $S$. Of
course, nonperturbative effects are responsible for the actual form of $f(S)$.

\epsfbox[-80 0 500 210]{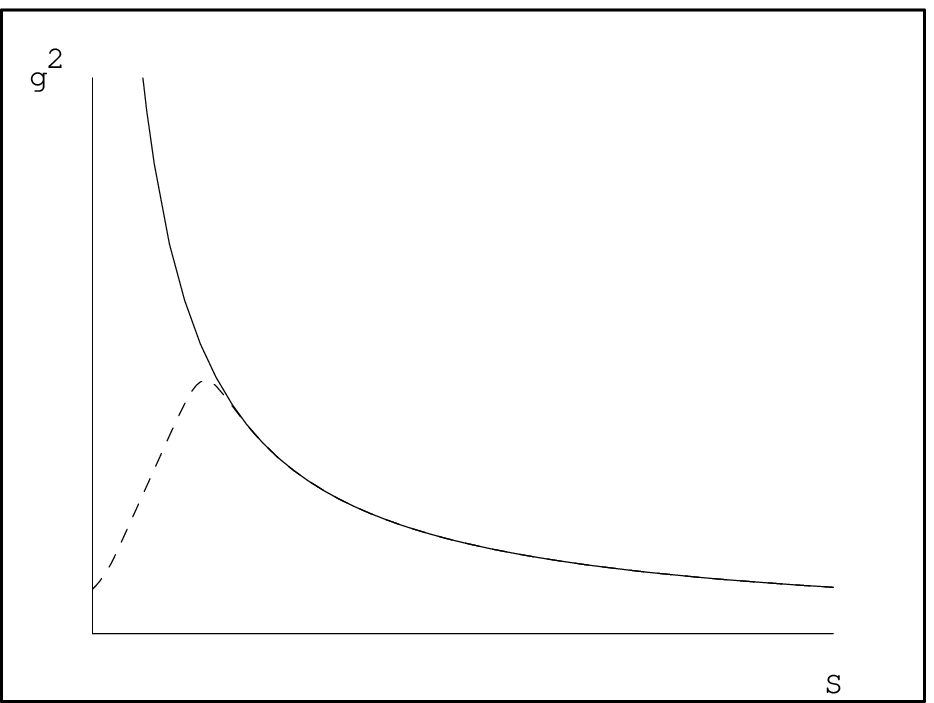}

{\small \em \noindent Fig. 1 - Coupling constant $g^2$ as
the function of $S$ in type-I models (dashed) vs $g^2$ given by $f=S$}

\vspace{0.3cm}

To examine the behaviour of the scalar potential in this approach, let us
consider a simple toy model, with chiral superfield $U=Y^3 = (\ll, \ldots )$ as
well as $S$ and $T$. We have to choose a specific example of a gauge kinetic
function which is invariant under the $S$-duality transformations. Different
choices are possible, the simplest is given by

\beq
f=\frac 1{2\pi} \ln(j(S)-744),
\eeq

\noindent $j(S)$ being the usual generator of modular invariant functions. This
function behaves like $S$ in the large $S$-limit. If we assume a type I-model
where the
gauge sector is closed under $S$-duality, then
we also have to assume that the gaugino condensate does not transform under
$S$-duality (because of the $f W^\alpha W_\alpha$-term in the
Lagrangian)\footnote{For type I-models it was shown in \cite{3} that one can
always redefine the gauge kinetic function and condensate in such a way that
this holds.}. Under these conditions an obvious candidate for the
superpotential is just the standard Veneziano-Yankielowicz superpotential
(extended to take into account the usual $T$-duality, which we assume to be
completely independent from $S$-duality) \cite{8}\cite{tdual}

\beq
W=Y^3 (f + 3b\ln \frac {Y\eta^2(T)}{\mu} +c).
\eeq

This is clearly invariant under $S$-duality. Therefore we then cannot take the
conventional form for the K\"ahler potential which would be given by

\beq
K= -\ln(S+\Sb)-3\ln(\TR),
\eeq

\noindent since it is not $S$-dual. To make it $S$-dual one could introduce an
additional $\ln |\eta(S)|^4$ term, giving e.g.

\beq
K=\ln(S+\Sb)-3\ln(\TR)-\ln |\eta(S)|^4.
\eeq

\noindent Because the only relevant quantity is

\beq
G=K+\ln |W|^2,
\eeq

\noindent we can as well put this new term (which is forced upon us because of
our demand
for symmetry) into the superpotential and take the canonical K\"ahler function
instead, which gives

\beq
K=\ln(S+\Sb)-3\ln(\TR),
\eeq
\beq
W=\frac{Y^3}{\eta^2(S)} (f + 3b\ln \frac {Y\eta^2(T)}{\mu} +c),
\eeq

\noindent where the remarkable similarity to the effective potential for
$T$-dual gaugino condensation $W=W_{inv}/\eta^6(T)$ can be seen more clearly.

This model exhibits a well defined minimum at $\vev S =1$, $\vev T=1.23$ and
$\vev Y\approx \mu$. Supersymmetry is broken with the dominant contribution
being $\vev{F_T}\approx \mu^3$. The cosmological constant is negative.

In contrast to earlier attempts \cite{ccm} this model fixes the
problem of the runaway dilaton and breaks supersymmetry with only a single
gaugino condensate. Previous models needed multiple gaugino condensates and (to
get realistic vevs for the dilaton) matter fields in complicated
representations. We feel that the concept of a nontrivial gauge kinetic
function derived (or constrained) by a symmetry is a much more natural way to
fix the dilaton and break supersymmetry, especially so because corrections to
$f=S$ are expected in any case. Earlier models which included $S$-duality in
different ways (both with and without gaugino condensates)
\cite{filq}\cite{hm}
were able to fix the vev of the dilaton but did not succeed in breaking
supersymmetry. An alternative mechanism to fix the vev of the dilaton has been
discussed in \cite{macross}.

Of course there are still some open questions not solved by this approach. The
first is the problem of having a vanishing cosmological constant. Whereas early
models of gaugino condensation often introduced {\em ad hoc} terms to guarantee
a vanishing vacuum energy, it has been seen to be notoriously difficult to get
this out of models based on string inspired supergravity. The only way out of
this problem so far has been to introduce a constant term into the
superpotential, parameterizing unknown effects. This approach does not even
work in any arbitrary model, but at least in our model the cosmological
constant can be made to vanish by adjusting such a constant.

Another question not addressed in this toy model is the mixing of $S$ and $T$
fields which happens at the one-loop level. It is still unknown whether one can
keep two independent dualities in this case. In a consistent interpretation our
toy model should describe an all-loop effective action. If it is considered to
be a theory at the tree-level then the
theory is not anomaly free. Introducing terms to cancel the anomaly
which arises because of demanding $S$-duality will then destroy $S$-duality. At
tree-level the theory therefore cannot be made anomaly free.

An additional interesting question concerns the vevs of the
auxiliary fields, i.e.  which field is responsible for
supersymmetry breakdown.
In all models considered so far (multiple gaugino condensates, matter,
S-duality) it has always been $F_T$ which dominates all the other auxiliary
fields. It has not been shown yet that this is indeed a generic feature. The
question is an important one, since the hierarchy of the vevs of the auxiliary
fields is mirrored in the structure of the soft SUSY breaking terms of the
MSSM \cite{brig}. We want to argue that there is at least no evidence for $F_T$
being generically large in comparison to $F_S$, because all of the models
constructed so far (including our toy model) are designed in such a way that
$\vev{F_S}=0$ by construction at the minimum (at least at tree-level for the
other models). In fact, if one extends our model with a constant in the
superpotential (see above), then $\vev{F_S}$ increases with the constant (but
does not become as large as $\vev{F_T}$).

Of course there are still some assumptions we made by considering this toy
model. We assumed that there is weak coupling in the large $S$ limit which is
an assumption because the nonperturbative effects are unknown (at tree-level it
can be calculated that $f=S$). In addition it is clear that the standard form
we take for the K\"ahler potential does not include nonperturbative effects and
thus could be valid only in the weak coupling approximation (this is of course
related to our choice of the superpotential). Of course an equally valid
assumption would be that nonperturbative effects destroy the calculable
tree-level behaviour even in the weak coupling region. The model of
ref. \cite{filq} could be re-interpreted in that sense (they do not consider
gaugino condensates and the gauge kinetic function, but their $S$-dual scalar
potential goes to infinity for $S\rightarrow \infty$). We choose not to make
this assumption, because it is equivalent to the statement that the whole
perturbative framework developed so far in string theory is wrong.
Again it should be emphasized here that the
$S$-duality considered is not a strong-weak coupling duality but a
weak-weak coupling duality. In type II-models one has a duality between strong
and weak coupling \cite{3}.

An additional problem could be the size of the gauge coupling constant. If
$f=S$ and $\vev S=1$ then the large value of the gauge coupling constant does
not fit the
low scale of gaugino condensation necessary for phenomenologically realistic
supersymmetry breaking ($10^{13}\,GeV$). However if $f=S$ only in the weak
coupling limit then one can have $\vev f >> 1$ and thus $g^2<< 1$ even in
the region $S=O(1)$. Therefore in our model $\vev S=1$ is consistent with the
demand for a small gauge coupling constant, whereas in models with $f=S$ a much
larger (and therefore more unnatural) $\vev S$ is needed.

To summarize we find that the choice of a nontrivial $f$-function (motivated by
a symmetry requirement) gives rise to a theory where supersymmetry breaking is
achieved by employing only a single gaugino condensate. The cosmological
constant turns out to be negative, but can be adjusted by a simple additional
constant in the superpotential. The vevs of all fields are at natural orders of
magnitude and due to the nontrivial gauge kinetic function the gauge coupling
constant can be made small enough to give a realistic picture.

\section*{Acknowledgments}

This work was supported by Deutsche Forschungsgemeinschaft under grant SFB-375
and the EC programs SC1-CT91-0729 and SC1-CT92-0789

\end{document}